\documentclass{ws-ijmpa}

\begin{document}

\markboth{Tao Zhu, Ji-Rong Ren, and Shu-Fan Mo} {Thermodynamics of
Friedmann Equation and Masslike Function in General Braneworld}

%%%%%%%%%%%%%%%%%%%%% Publisher's Area please ignore %%%%%%%%%%%%%%%
%
\catchline{}{}{}{}{}
%
%%%%%%%%%%%%%%%%%%%%%%%%%%%%%%%%%%%%%%%%%%%%%%%%%%%%%%%%%%%%%%%%%%%%

\title{Thermodynamics of Friedmann Equation and Masslike Function in General Braneworld}

\author{Tao Zhu, Ji-Rong Ren, and Shu-Fan Mo}
\address{Institute of Theoretical Physics, Lanzhou University,
Lanzhou 730000, China\\
zhut05@lzu.cn, renjr@lzu.edu.cn, meshf07@lzu.cn}

\maketitle

\begin{history}
\received{Day Month Year}
\revised{Day Month Year}
\end{history}

\begin{abstract}
Using the generalized procedure proposed by \emph{Wu et al}\cite{wu}
recently, we construct the first law of thermodynamics on apparent
horizon in a general braneworld model with curvature correction
terms on the brane and in the bulk, respectively. The explicit
entropy formulary of apparent horizon in the general braneworld is
worked out. We also discuss the masslike function which associated
with a new type first law of thermodynamics of the general
braneworld in detail. We analyze the difference between the
conventional thermodynamics and the new type thermodynamics on
apparent horizon. At last, the discussions about the physical
meanings of the masslike function have also been given.

\end{abstract}

\ccode{PACS numbers: 98.80.Cq, 11.10.Kk, 11.25.Wx}

\section{Introduction}

The four thermodynamics laws of black hole, which were originally
derived from the classical Einstein Equation, provide deep insight
into the connection between thermodynamics and Einstein
Equation\cite{Black Hole}\cdash\cite{Black Hole4}. In
\emph{Jacobson}'s pioneer paper\cite{Jacobson}, this connection has
been extended into a general spacetime. In \emph{Jacobian}'s set-up,
the Einstein equation can be derived from the proportionality of
entropy to the horizon area, together with the Clausius relation
$\delta Q=TdS$. Here $\delta Q$ and $T$ are the energy flux and
Unruh temperature detected by an accelerated observer just inside
the local Rindler causal horizons through spacetime point. From the
viewpoint of thermodynamics, the Einstein equation can be regarded
as the equation of state of spacetime. Since \emph{Jacobson}'s work,
many physicists have extended the connection between thermodynamics
and gravity beyond the Einstein gravity, including the so-called
$f(R)$ gravity\cite{fr1}\cdash\cite{fr-st} and the scalar-tensor
gravity\cite{fr-st,love-scalar-tensor}. Recently, the connection
between thermodynamics of apparent horizon and Friedmann Equation in
Friedmann-Robertson-Walker (FRW) universe has been
shown\cite{fr2,fr-st}\cdash\cite{love-scalar-tensor}. This
connection has also been extended to braneworld cosmology, for
related discussions see Refs.~\refcite{brane1}--\refcite{brane-cai}.
On the other hand, the thermodynamics of FRW universe has also been
studied in term of holographic principle\cite{cft/frw,cft/frw-1}.

More recently, \emph{Eling et al}\cite{fr1} found that it is
impossible to derive the field equation of $f(R)$ gravity from the
Clausius relation $\delta Q=TdS$, in terms of equilibrium
thermodynamics. In order to get the field equation in $f(R)$
gravity, an entropy production term has to be added to the Clausius
relation which is then changed to $\delta Q=TdS+Td_iS$. Similar
cases have also occurred in the scalar-tensor
theory\cite{love-scalar-tensor}. The thermodynamics behaviors in
$f(R)$ gravity and scalar-tensor gravity show that we have to treat
these system with non-equilibrium thermodynamics, which are
different with the equilibrium thermodynamics in Einstein gravity.
In Ref.~\refcite{masslike}, by introducing a masslike function, the
authors showed that the equilibrium thermodynamics on apparent
horizon of FRW universe can exist for more general theories of
gravity, even including $f(R)$ gravity, scalar-tensor gravity, and
Brans-Dicke theory. The masslike function provides a new type first
law of thermodynamics on apparent horizon of FRW universe, which is
obvious different with the conventional thermodynamical treatment of
FRW spacetime. Then \emph{Wu et al}\cite{wu} proposed a general
procedure to construct the first law of thermodynamics on apparent
horizon in generalized gravity theories, and discussed a more
general formulary for the masslike function. The validity of their
procedure has been shown in $f(R)$ gravity, Lovelock gravity,
scalar-tensor gravity, and also the Randal-Sundrum braneworld with
bulk matter.

However, the universality of the procedure in a more general
braneworld model has not been discussed. In this paper, we employ
this procedure to study the connection between thermodynamics and
the general braneworld model with correction terms, such as a $4D$
scalar curvature from the induced gravity on the brane, and a $5D$
Gauss-Bonnet curvature term. The connection between thermodynamics
and this general braneworld model have also been investigated in
Ref.~\refcite{brane-cai}. They have derived the entropy expression
of the apparent horizon even though the exact analytic black hole
solution is absent so far. We expect that the entropy formulary
derived from the general procedure that proposed by \emph{Wu et al}
should be consistent with the entropy expression in
Ref.~\refcite{brane-cai}.

It is also interesting to explore whether and how the connection
between thermodynamics associated with the masslike function and
gravity theories be generalized to the braneworld scenarios. There
are two motivations encourage us to address this issue. First, the
new type first law of thermodynamics on apparent horizon which
related to the masslike function is a geometric relation,
\begin{eqnarray}
dE=TdS,\label{gr}
\end{eqnarray}
where $T$ and $S$ are both geometric quantities. The energy flow
through the apparent horizon is defined by a masslike function. In a
general braneworld, the curvature correction terms on the brane and
in the bulk must affect the energy flow crossing the apparent
horizon on the brane. In this case, does this geometric relation
(\ref{gr}) also hold when the curvature correction terms exist?
Whether and how do the contribution of the curvature correction
terms enter the expression of the masslike function? These questions
need to be answered. Second, in the braneworld scenario, the gravity
on the brane is not the Einstein gravity, the extra dimension effect
on the brane may also affect the masslike function. These
non-trivial contributions to the masslike function provide an
abundant physical meanings of the masslike function, this may give
some clues to explore the physical meanings of the masslike function
and the universal geometric relation (\ref{gr}). In this paper, we
will discuss the connection between thermodynamics and the
braneworld model. The new type first law of thermodynamics related
to the masslike function will be investigated in detail.

The paper is organized as follows. In Section II, we give a brief
introduction of the general procedure proposed by \emph{Wu et
al}\cite{wu} by generalizing it to the case of a FRW universe with
any spatial curvature. In section III, we consider a FRW universe on
the brane, and construct the first law of thermodynamics on apparent
horizon and calculate the entropy of apparent horizon in the general
braneworld. In Section IV, we investigate the universality of the
geometric relation (\ref{gr}) in the braneworld. The masslike
function has been worked out. The physical meanings of the masslike
function have also been explored. We end this paper with conclusion
in Section V.

\section{First Law of Thermodynamics of Friedmann Equation on Apparent Horizon}
In this section, we will give a brief introduction of the general
procedure that proposed by \emph{Wu et al} by generalizing it to the
case of a FRW universe with any spatial curvature. Let us start with
an $(n+1)$-dimensional homogenous and isotropic FRW universe, whose
metric is
\begin{eqnarray}
ds^2=h_{ab}dx^adx^b+\tilde{r}^2d\Omega^{2}_{n-1},
\end{eqnarray}
where $x^0=t, x^1=r$, $\tilde{r}=ar$ is the radius of the sphere,
$a$ is the scale factor, and $d\Omega^{2}_{n-1}$ is the $(n -
1)$-dimensional sphere element. Here
$h_{ab}=$diag$(-1,a^2/(1-kr^2))$ is the $2$-dimensional metric, in
which $k$ is the spatial curvature constant. The dynamical apparent
horizon, a marginally trapped surface with vanishing expansion, is
defined by $h^{ab}\partial_a\tilde{r}\partial_b\tilde{r}=0$, from
this relation the radius of the apparent horizon reads
\begin{eqnarray}
\tilde{r}_A=\frac{1}{\sqrt{H^2+k/a^2}}=\frac{1}{\mathcal{H}},\label{Horizon}
\end{eqnarray}
where $H$ denotes the Hubble parameter, $H=\dot{a}/a$, and for
convenient, we define $\mathcal{H}^2=H^2+k/a^2$. Here we set the
dots represent derivatives with respect to the cosmic time $t=x^0$.
The associated temperature on the apparent horizon can be defined as
\begin{eqnarray}
T=\frac{1}{2\pi \tilde{r}_A}.
\end{eqnarray}
In Einstein gravity, the entropy is proportional to the horizon area
\begin{eqnarray}
S_E=\frac{A}{4G},\label{entropy-area}
\end{eqnarray}
where the horizon area $A=n\Omega_{n}\tilde{r}_{A}^{n-1}$, thus we
have the fundamental relation
\begin{equation}
\delta Q\equiv
TdS_{E}=\frac{n(n-1)V\tilde{r}_{A}^{-3}d\tilde{r}_{A}}{8\pi G},
\end{equation}
where $V=\Omega_{n}\tilde{r}_{A}^{n}$ is the volume in the horizon.
Using the relation (\ref{Horizon}), we can obtain
\begin{equation}
TdS_{E}=\frac{-n(n-1)V}{16\pi G}\frac{d\mathcal{H}^{2}}{dt}dt,
\label{dr1}
\end{equation}
where $T, S,$ and $\mathcal{H}$ are pure geometric quantities, this
implies that the above relation (\ref{dr1}) is a pure geometric
relation.

For Einstein gravity theories, one can write the Friedmann equations
in the form
\begin{equation}
\mathcal{H}^2=H^{2}+\frac{k}{a^2}=\frac{16\pi G}{n(n-1)}\rho _{eff}.
\label{H21}
\end{equation}
Though we do not know the exact form of $\rho _{eff}$ (and
$p_{eff}$), we know that there must be ordinary matter density $\rho
$ in $\rho _{eff}$ and also other quantities $\rho _{i}$, such as
matter or energy components besides the ordinary matter. For all
gravity theories, the Friedmann equation can be expressed in a
generalized form
\begin{equation}
f(\mathcal{H}^2, \rho, \;\rho _{1},\cdots \rho _{i},\cdots
)=0.\label{friedmann}
\end{equation}
It is obvious shown that $\mathcal{H}^2$ is not only dependent on
ordinary matter density $\rho $, but also other quantities $\rho
_{i}$, i.e.,
\begin{equation}
\mathcal{H}^{2}=\mathcal{H}^{2}(\rho ,\;\rho _{1},\cdots \rho
_{i},\cdots ).
\end{equation}
In the general braneworld models, we will show in the next section
that $\mathcal{H}^2$ only dependent on the ordinary matter density
$\rho $, therefore, in the following discussions, we will restrict
to consider this case only for simplicity. Then the relation
(\ref{dr1}) can be changed to
\begin{equation}
TdS_{E}=\frac{-n(n-1)V}{16\pi G}\frac{\partial
\mathcal{H}^{2}}{\partial \rho }\dot{\rho}dt. \label{dr2}
\end{equation}
The expression of $\frac{\partial \mathcal{H}^{2}}{\partial \rho }$
can be got from the Friedmann Equation (\ref{friedmann}). To
construct the first law of thermodynamics $dE=TdS$, we need to know
the energy flux $dE$ and entropy $S$. In the general gravity theory,
they are not specified. The energy flux of ordinary matter can be
expressed as $dE=V \dot{\rho}dt$. Multiplying $\frac{16\pi
G}{n(n-1)}(\frac{\partial \mathcal{H}^{2}}{\partial \rho })^{-1}$ on
both sides of (\ref{dr2}), we have
\begin{equation}
\frac{16\pi G}{n(n-1)}\bigg(\frac{\partial \mathcal{H}^{2}}{\partial
\rho }\bigg)^{-1} TdS_{E}=-V\dot{\rho}dt. \label{dr3}
\end{equation}
In the general case, the conservation of the ordinary matter density
can be written as
\begin{equation}
\dot{\rho}+nH(\rho +p)=0.  \label{Con2}
\end{equation}
Substituting $\dot{\rho}$ into Eq.(\ref{dr3}), we can get
\begin{equation}
T\frac{16\pi G}{n(n-1)}\left(\frac{\partial
\mathcal{H}^{2}}{\partial \rho }\right)^{-1} dS_{E}=nVH(\rho +p)dt.
\label{dr4}
\end{equation}
The entropy form can be got by integrating (\ref{dr4}). If there is
just ordinary matter $\rho $ in the space, $\frac{\partial
\mathcal{H}^{2}}{\partial \rho }$ can be rewritten as a function of
$\tilde{r}_{A}$. Then the entropy can be obtained by the integration
\begin{equation}
S=\int \frac{16\pi G}{n(n-1)}\left(\frac{\partial
\mathcal{H}^{2}}{\partial\rho }\right)^{-1}dS_{E}=\int 4\pi
\tilde{r}_{A}^{n-2}\Omega _{n}\left(\frac{\partial
\mathcal{H}^{2}}{\partial\rho }\right)^{-1}d\tilde{r}_{A},
\label{S0}
\end{equation}
so the entropy formulary is obviously dependent on $\frac{\partial
\mathcal{H}^{2}}{\partial\rho }$. This is the crucial result which
can be used to determine the exact entropy formulary for general
braneworld models. Then the relation (\ref{dr4}) can be written as
\begin{equation}
TdS=dE,  \label{first law1}
\end{equation}
where $dE=V\dot{\rho}dt=nVH(\rho +p)dt$. It is the first law of
thermodynamics for the gravity theories with only freedom $\rho $ in
the first Friedmann equation.

When $\mathcal{H}^2$ is not only dependent on ordinary matter
density $\rho $, such as in $f(R)$ gravity, scalar-tensor gravity,
and also Brans-Dicke Theory, the general expression of the first law
of thermodynamics in the Friedmann equation reads
\begin{equation}
TdS+Td_iS=dE,
\end{equation}
where $d_iS$ is interesting since it relates to the entropy
production in the non-equilibrium thermodynamics.

\section{Thermodynamics of Friedmann Equation and Entropy Formulary in General Braneworlds}
In this section, we will use the above procedure to investigate the
thermodynamics properties of Friedmann Equation and the Entropy
Formulary in General Braneworld. We consider a $3$-brane embedded in
a $4+1$-dimensional space-time. For convenience and without loss of
generality we choose the extra dimension along the coordinates $y$
such that the brane is located at $y = 0$. Objects corresponding to
the brane are written with a tilde to be distinguished from $5D$
objects. We begin with the
action\cite{brane-cai,brane-jhep}\cdash\cite{brane-jhep-2}
\begin{eqnarray}
S&=&\frac{1}{2\kappa_5}\int dx^5\sqrt{-g}(R-2\Lambda+\alpha\mathcal
{L}_{GB})\nonumber\\
&+&\frac{1}{2\kappa_4}\int dx^4\sqrt{-\tilde{g}}\tilde{R}+\int
dx^4\sqrt{-\tilde{g}}(\mathcal {L}_{m}-2\lambda),
\end{eqnarray}
where $\Lambda<0$ is the bulk cosmological constant and $\mathcal
{L}_{GB}=R^2-4R^{AB}R_{AB}+R^{ABCD}R_{ABCD}$ is the Gauss-Bonnet
correction term. $g$ is the bulk metric and $R$, $R_{AB}$, and
$R_{ABCD}$ are the curvature scalar, Ricci, and Riemann tensors,
respectively. $\kappa_4$ and $\kappa_5$ are the gravitational
constants on the brane and in the bulk, respectively. $\mathcal
{L}_m$ is the Lagrangian density of the brane matter fields, and
$\lambda$ is the brane tension (or the brane cosmological constant).
For convenience, we assume that the brane cosmological constant is
zero. We assume that there are no sources in the bulk other than
$\Lambda$ and redefine $\kappa^2_4=8\pi G_4$, $\kappa^2_5=8\pi G_5$.

We consider homogeneous and isotropic brane at fixed coordinate
position $y=0$ in the bulk, the bulk metric is described by
\begin{equation}
ds^2=-N^2(t,y)dt^2+A^2(t,y)\gamma_{ij}dx^idx^j+B^2(t,y)dy^2,
\end{equation}
where $\gamma_{ij}$ is a constant curvature three-metric, with
curvature index $k$. For this metric, the generalized Friedmann
equation on the brane has been obtained in
Refs.~\refcite{brane-cai,brane-jhep}--\refcite{brane-jhep-2},
\begin{equation}
-\frac{2\kappa_4^2}{\kappa_5^2}\left[1+\frac{8}{3}\alpha(\mathcal{H}^2+\frac{\Phi_0}{2})\right]
(\mathcal{H}^2-\Phi_0)^{1/2}=-\frac{\kappa_4^2}{3}\rho+\mathcal{H}^2,\label{friedmann2}
\end{equation}
in which $\Phi$ is defined as
\begin{equation}
\Phi=\frac{1}{N^2}\frac{\dot{A}^2}{A^2}-\frac{1}{b^2}\frac{A^{'2}}{A^2}+\frac{k}{A^2},
\end{equation}
and $\Phi_0=\Phi(t,0)$. In order to compare our discussion with the
result obtained in\cite{brane-cai}, we use the same assumption that
there is no black hole in the bulk and so
$\Phi_0=\frac{1}{4\alpha}(-1+\sqrt{1+\frac{4\alpha\Lambda}{3}})=const$.

Noticing now that $k$, $\kappa_4$, $\kappa_5$ and $\Phi_0$ are all
constant, it is obvious that the Friedmann equation
(\ref{friedmann2}) is consistent with the general form
\begin{equation}
f(\mathcal{H}^2,\rho)=\frac{2\kappa_4^2}{\kappa_5^2}\left[1+\frac{8}{3}\alpha(\mathcal{H}^2+\frac{\Phi_0}{2})\right]
(\mathcal{H}^2-\Phi_0)^{1/2}-\frac{\kappa_4^2}{3}\rho+\mathcal{H}^2=0.\label{friedmann3}
\end{equation}
It is obvious that the $\mathcal{H}^2$ is only dependent on $\rho$.
In order to search the expression of $\frac{d\mathcal{H}^2}{d\rho}$,
we reexpress Eq.(\ref{friedmann3}) as
\begin{eqnarray}
f&=&\frac{2\kappa_4^2+8\kappa_4^2\alpha\Phi_0}{\kappa_5^2}(\mathcal{H}^2-\Phi_0)^{1/2}+
(\mathcal{H}^2-\Phi_0)\nonumber\\
&+&\frac{16\kappa_4^2\alpha}{3\kappa_5^2}(\mathcal{H}^2-\Phi_0)^{3/2}-\frac{\kappa_4^2}{3}\rho
+\Phi_0=0.\label{friedmann4}
\end{eqnarray}
Operate with $\frac{d}{d\rho}$ on the above equation, after several
steps of simple calculation, we get
\begin{equation}
\left(\frac{d\mathcal{H}^2}{d\rho}\right)^{-1}=\frac{3}{8\pi G_4}+
\frac{3}{8\pi G_5}\frac{\tilde{r}_A}{\sqrt{1-\Phi_0
\tilde{r}^2_A}}+\frac{3\alpha}{2\pi
G_5}\frac{2-\Phi_0\tilde{r}^2_A}{\sqrt{1-\Phi_0
\tilde{r}^2_A}}\frac{1}{\tilde{r}_A}.\label{friedmann5}
\end{equation}
Since $\mathcal{H}^2$ is only dependent on $\rho$ in the Friedmann
equation, noticing that $n=3$ and $G=G_4$ on the $3$-brane and
making use of the entropy expression (\ref{S0}) and
Eq.(\ref{friedmann5}), we obtain the entropy associated with the
apparent horizon on the brane as
\begin{equation}
S=\frac{3\Omega_3}{2G_4}\int\tilde{r}_Ad\tilde{r}_A+
\frac{3\Omega_3}{2G_5}\int\frac{\tilde{r}^2_Ad\tilde{r}_A}{\sqrt{1-\Phi_0
\tilde{r}^2_A}}+\frac{6\alpha\Omega_3}{G_5}\int\frac{2-\Phi_0\tilde{r}^2_A}{\sqrt{1-\Phi_0
\tilde{r}^2_A}}d\tilde{r}_A.\label{entropy}
\end{equation}
Integrating the above expression, the explicit form of the entropy
can be obtained as
\begin{eqnarray}
S&=&\frac{3\Omega_3\tilde{r}^2_A}{4G_4}+
\frac{2\Omega_3\tilde{r}^3_A}{4G_5}~{}_2F_{1}(\frac{3}{2},\frac{1}{2},\frac{5}{2},\Phi_0
\tilde{r}^2_A)\nonumber\\
&+&\frac{6\alpha\Omega_3\tilde{r}^3_A}{G_5}\left[\Phi_0~{}_2F_{1}(\frac{3}{2},\frac{1}{2},\frac{5}{2},\Phi_0
\tilde{r}^2_A)+\frac{\sqrt{1-\Phi_0
\tilde{r}^2_A}}{\tilde{r}^2_A}\right],\label{entropy2}
\end{eqnarray}
where ${}_2F_{1}(\frac{3}{2},\frac{1}{2},\frac{5}{2},\Phi_0
\tilde{r}^2_A)$ is a hypergeometric function. This expression is
exactly consistent with the entropy formulary obtained by
\emph{Sheykhi et al}\cite{brane-cai}. The corresponding first law of
thermodynamics (\ref{first law1}) reads
\begin{equation}
TdS=dE=3VH(\rho+p)dt.\label{first law2}
\end{equation}
This is just the Clausius relation in the version of black hole
thermodynamics. From (\ref{first law2}), we can see clearly that
there is no additional entropy production term $d_iS$, this implies
that the thermodynamics we treated in the general braneworld is
equilibrium thermodynamics.

Although the entropy formulary we obtained is the same as that
obtained in Ref.~\refcite{brane-cai}, the expressions of the first
law of thermodynamics on apparent horizon are different. In
Ref.~\refcite{brane-cai}, \emph{Sheykhi et al} obtained the entropy
formulary by applying the first law of thermodynamics, $TdS+WdV=dE$,
to the apparent horizon of a FRW universe on the brane, while in
this paper, the first law of thermodynamics we applied is $TdS=dE$.
We would like to point out here that this difference is not worth to
worry, because the result in Ref.~\refcite{brane-cai} is consistent
with the one in this paper. As pointed out in
Ref.~\refcite{gauss-love2}, this difference is due to the different
definitions of $dE$. In Ref.~\refcite{brane-cai}, $E$ is the matter
energy inside the apparent horizon, $E=\Omega_n \tilde{r}_A^n \rho$.
The change of energy $dE$ inside the apparent horizon is
\begin{equation}
dE=n\Omega_n \tilde{r}_A^{n-1}\rho d\tilde{r}_A-n\Omega_n
\tilde{r}_A^{n}(\rho+p)Hdt.
\end{equation}
In this paper, the definition of $dE$ is
\begin{equation}
dE=n\Omega_n\tilde{r}_A^n(\rho+p)Hdt,
\end{equation}
while the term of volume change is absent in our consideration.

So far, we have constructed the first law of thermodynamics of
Friedmann equation on apparent horizon in a general braneworld. As
expected, the entropy formulary of apparent horizon obtained in this
section is consistent with the entropy formulary obtained by
\emph{Sheykhi et al}\cite{brane-cai}. Now, we give some remarks
about above discussions in order: (i) As pointed out in
Ref.~\refcite{brane-cai}, the physical meaning of the entropy
formulary (\ref{entropy2}) is obvious. The first term in
(\ref{entropy2}) is Bekenstein- Hawking entropy on the brane, the
second term obeys the 5-dimensional area formula in the bulk and the
third term come off the contribution of the Gauss-Bonnet correction
term. (ii) The Eq.(\ref{entropy2}) is a very general entropy
formulary in braneworld, it can reduce to the entropy of several
special braneworld models\cite{brane1,brane2,brane-cai}. Such as the
Dvali-Gabadadze-Porrati (DGP) braneworld is the limiting case when
$\alpha=0$, the Randall-Sundrum (RS) II braneworld in the limit
$\kappa_4\rightarrow\infty$ and $\alpha=0$, the pure Gauss-Bonnet
braneworld is the case with $\kappa_4\rightarrow\infty$.

\section{Masslike Function in General Braneworlds}
In this section, we will begin to study the first law of
thermodynamics associated with the masslike function and search for
the expression of the masslike function in braneworld. As shown in
Ref.~\refcite{masslike}, the masslike function in $(3 +
1)$-dimensional Einstein gravity reads
\begin{equation}
M=\frac{\tilde{r}}{2G}(1+h^{ab}\partial_a\tilde{r}\partial_b\tilde{r}).
\end{equation}
Using this masslike function, the first law of Einstein gravity
reads
\begin{equation}
TdS_E=k^a\partial_aMdt=dE,
\end{equation}
where $k^a=(-1,Hr)$ is null (approximate) generator of the apparent
horizon. The above expression plays a important role in determining
the exact expression of the masslike function in modified gravity.
As pointed out in Ref.~\refcite{wu}, this masslike function can be
insteaded by a more generalized form. Using Eq.(\ref{dr4}), the
masslike function satisfied
\begin{equation}
\frac{16\pi G}{n(n-1)}(\frac{\partial \mathcal{H}^{2}}{\partial \rho
})^{-1}TdS_E=\frac{16\pi G}{n(n-1)}(\frac{\partial
\mathcal{H}^{2}}{\partial \rho
})^{-1}k^a\partial_a(M+f_1)dt=k^a\partial_a\tilde{M}dt,\label{mass1}
\end{equation}
where $M$ is the $(n+1)$-dimensional masslike function, which reads
\begin{equation}
M=\frac{n(n-1)\Omega_n\tilde{r}^{n-2}}{16\pi
G}(1+h^{ab}\partial_a\tilde{r}\partial_b\tilde{r}),
\end{equation}
and $f_1$ and also $f_2$ (which will be defined bellow) are
arbitrary functions satisfying
\begin{equation}
k^a\partial_af_i=0~~~~~(i=1,~2)
\end{equation}
on the apparent horizon. From Eq.(\ref{mass1}), we get
\begin{equation}
\tilde{M}=\frac{16\pi G}{n(n-1)}(\frac{\partial
\mathcal{H}^{2}}{\partial \rho })^{-1}(M+f_1)+f_2.\label{m}
\end{equation}
Noticing that $n=3$ and $G=G_4$ on the brane, substituting
Eq.(\ref{friedmann5}) into Eq.(\ref{m}), the above formulary gives
the exact expression of the masslike function $\tilde{M}$ in the
braneworld,
\begin{eqnarray}
\tilde{M}&=&\left(1+
\frac{G_4}{G_5}\frac{\tilde{r}_A}{\sqrt{1-\Phi_0
\tilde{r}^2_A}}+\frac{4\alpha G_4}{
G_5}\frac{2-\Phi_0\tilde{r}^2_A}{\sqrt{1-\Phi_0
\tilde{r}^2_A}}\frac{1}{\tilde{r}_A}\right)\nonumber\\
&\times&\left(\frac{3\Omega_3\tilde{r}}{8\pi
G_4}(1+h^{ab}\partial_a\tilde{r}\partial_b\tilde{r})+f_1\right)+f_2.
\end{eqnarray}
Using this masslike function, the first law of the general
braneworld now reads
\begin{equation}
TdS=k^a\partial_a\tilde{M}dt=dE.\label{first law3}
\end{equation}
This result exactly has the same form as the one that we have given
in the previous section.

In addition, we would like to point out that,  although
Eq.(\ref{first law2}) and (\ref{first law3}) have the same form, the
first law of thermodynamics expressed in Eq.(\ref{first law3}) is a
new type thermodynamics and the corresponding physical meanings are
different. First, the corresponding definition of the energy flow
$dE$ through apparent horizon are different. Because of the masslike
function, the energy flow $dE$ defined in (\ref{first law3})
includes the contribution of gravitational field such as the
Gauss-Boned term and bulk contribution in the general braneworld, in
addition to the matter field on the brane. But in previous section,
the energy flow $dE$ is defined as the matter field energy crossing
the apparent horizon within an infinitesimal of time $dt$. Second,
the universality of Eq.(\ref{first law2}) and (\ref{first law3}) are
obvious different. In $f(R)$ gravity and scalar-tensor
theory\cite{fr2,fr-st,love-scalar-tensor}, the equilibrium
thermodynamics relation Eq.(\ref{first law2}) does not hold on
apparent horizon for Friedmann equation. In order to construct the
thermodynamics of apparent horizon for Friedmann equation, one has
to modify the thermodynamics relation (\ref{first law2}) to
nonequilibrium case by adding an entropy production term. But for
the relation Eq.(\ref{first law3}) associated with the masslike
function, its validity has been verified in various gravity
theory\cite{masslike,wu}, including $f(R)$ gravity, scalar-tensor
theory, Gauss-Bonnet gravity and more general Loveloke Gravity. And
in this section, it also hold in the general braneworld model.

The masslike function $\tilde{M}$ obtained above plays a important
role in the thermodynamic description of the gravitational dynamics
and determines the energy flows passing through the horizon. For a
variety of theories of gravity, the masslike function reduces to the
Misner-Sharp mass at the apparent horizon. Therefore, the
investigation of this mass-like function may shed lights on the
Misner-Sharp mass of the braneworld. In order to conveniently
discuss the physical meanings of the masslike function, we set
$f_1=0$ and $f_2=0$, and notice that
$h^{ab}\partial_a\tilde{r}\partial_b\tilde{r}=0$ on the apparent
horizon, the masslike function $\tilde{M}$ reduces to
\begin{equation}
\tilde{M}=\frac{3\Omega_3\tilde{r}_A}{8\pi G_4}+
\frac{3\Omega_3\tilde{r}^2_A}{8\pi G_5}\frac{1}{\sqrt{1-\Phi_0
\tilde{r}^2_A}}+\frac{3\Omega_3\alpha}{2\pi
G_5}\frac{2-\Phi_0\tilde{r}^2_A}{\sqrt{1-\Phi_0
\tilde{r}^2_A}}.\label{mass}
\end{equation}
It is obvious that this masslike function contains the contributions
of the extra dimension and the Gauss-Bonnet correction term in the
bulk. This means that the energy flows passing through the horizon
on the brane may have some non-trivial connection with the extra
dimension and the Guass-Bonnet curvature correction terms in the
bulk. This non-trivial connection also has some interesting effects
on the accelerated expansion of the
universe\cite{energy-change}\cdash\cite{wu3}.

In the context of dynamical black holes, the dynamics of the black
hole spacetime can be described by its quasi-local first law of
thermodynamics, this thermodynamics associated with a quasi-local
mass which determines the energy (or mass) of the trapping horizon
of black holes\cite{ql1}\cdash\cite{ql3-1}. In general relativity,
the quasi-local mass usually be selected as the Misner-Sharp
mass\cite{ql1}\cdash\cite{ql2}. Such quasi-local mass has also been
generalized to the Einstein-Gauss-Bonnet gravity by Hideki Maeda and
Asato Nozawa\cite{ql3,ql3-1}. In FRW spacetime, the apparent horizon
is an important trapping horizon, at this level, its thermodynamics
coincides with the quasi-local thermodynamics of dynamical trapping
horizons. For the models of braneworld considered in this paper, the
new type first law of thermodynamics associated with masslike
function is just the quasi-local first law of thermodynamics. This
implies that the above masslike function (\ref{mass}) should
coincides with the quasi-local mass of apparent horizon, which can
be regarded as the generalized Misner-Sharp mass in braneworlds.
Therefore, the thermodynamics of apparent horizon can provide an
approach to investigate the properties of Misner-Sharp mass in the
braneworlds.

 From Eq.(\ref{mass}), we
can obtain the masslike function of several special braneworld
models:

(i). In the limit $\alpha\rightarrow 0$, Eq.(\ref{mass}) reduces to
the masslike function on the apparent horizon in the warped DGP
braneworld embedded in an $AdS_5$ bulk
\begin{equation}
\tilde{M}=\frac{3\Omega_3\tilde{r}_A}{8\pi G_4}+
\frac{3\Omega_3\tilde{r}^2_A}{8\pi G_5}\frac{1}{\sqrt{1-\Phi_0
\tilde{r}^2_A}}.
\end{equation}

(ii). When in the limit $\alpha\rightarrow 0$ and $\Phi_0\rightarrow
0$, Eq.(\ref{mass}) reduces to the masslike function on the apparent
horizon in pure DGP braneworld with a Minkowskian bulk
\begin{equation}
\tilde{M}=\frac{3\Omega_3\tilde{r}_A}{8\pi G_4}+
\frac{3\Omega_3\tilde{r}^2_A}{8\pi G_5}.
\end{equation}

(iii). In the limit $\alpha\rightarrow 0$ and
$G_4\rightarrow\infty$, while keeping $G_5$ finite, the first and
the last terms in Eq.(\ref{mass}) vanish and we obtain the masslike
function on the apparent horizon in the RS II braneworld
\begin{equation}
\tilde{M}=\frac{3\Omega_3\tilde{r}^2_A}{8\pi
G_5}\frac{1}{\sqrt{1-\Phi_0 \tilde{r}^2_A}}.\label{massz}
\end{equation}

(iv). Finally, keeping $\alpha$ finite, and in the limit
$G_4\rightarrow\infty$ and $\Phi_0\rightarrow 0$, we obtain the
masslike function on the apparent horizon in the Gauss-Bonnet
braneworld with a Minkowskian bulk
\begin{equation}
\tilde{M}=\frac{3\Omega_3\tilde{r}^2_A}{8\pi
G_5}(1+\frac{24\alpha}{\tilde{r}^2_A}).\label{massy}
\end{equation}
Although, the connections between these masslike functions and
Misner-Sharp mass are not very clear, it is predictable that these
masslike functions may play an important role in investigating the
Misner-Sharp mass in the braneworld.

\section{Conclusion and Discussions}
In this paper we have studied the thermodynamics of the apparent
horizon of FRW Universe in a general braneworld model with curvature
correction terms on the brane and in the bulk, respectively. Using
the general procedure developed by \emph{Wu et al}, we have
constructed the first law of thermodynamics on apparent horizon of
FRW Universe and obtained the exact entropy formulary of apparent
horizon in the braneworld. As expected, the entropy formulary we
obtained is the same as the one obtained by \emph{Sheykhi et
al}\cite{brane-cai}.

We have also studied the first law of thermodynamics of apparent
horizon associated with the masslike function in the braneworld.
This is a new type first law of apparent horizon and it is a
universality result in more generalized gravity theories. Its
validity in the braneworld has been verified. The difference between
this new type first law of thermodynamics and the conventional first
law of thermodynamics in the braneworld have also been discussed. We
have also calculated the exact expression of the masslike function
in the braneworld. As expected, the masslike function contains the
contributions of the extra dimension and the Guass-Bonnet curvature
correction terms in the bulk. This implies that the energy flow
crossing the apparent horizon on the brane should contain the
contributions from bulk and curvature correction terms. The physical
meanings of the masslike function have been discussed in the context
of quasi-local thermodynamics in the dynamical horizon. As like in
the Einstein gravity, $f(R)$ gravity, and the LoveLock gravity the
masslike function reduces to the Misner-Sharp mass on apparent
horizon, we concluded that the masslike function should also reduce
to the generalized Misner-Sharp mass on apparent horizon in the
braneworld.

We noticed that in our construction of first law of thermodynamics
in Section III, the energy conservation on the brane is assumed.
This means that there is no interaction between extra dimension and
the matter on the brane. When the bulk matter is assumed, the energy
conservation on the brane may not hold. It is of great interest to
extend the thermodynamics to the braneworld model with bulk matter
content.

\section*{Acknowledgments}

The author \emph{Tao Zhu} thanks \emph{Dr. Mariam Bouhmadi-Lopez}
for bringing our attention to
Refs.~\refcite{pham}--\refcite{DGP-LOPEZ} and thanks \emph{Dr.
Hideki Maeda} for his useful comments and discussion about this
work. We also thank \emph{Dr. Ming-Fan Li} for his works in
improving the English Writing of this manuscript. This work was
supported by the National Natural Science Foundation of China
(No.10275030) and the Cuiying Programme of Lanzhou University
(225000-582404).

%\begin{thebibliography}{000} %for 3 digits
%\begin{thebibliography}{00}  %for 2 digits

\end{document}